\documentclass[12pt]{article}
\usepackage[pctex32]{graphics}
\begin{document}
~
~
\vspace{1cm}
\begin{center} {\Large \bf  Multi-spin String Solutions in Magnetic-flux Deformed $AdS_n \times S^m$ Spacetime}
                                                  
\vspace{1cm}

                      Wung-Hong Huang\\
                       Department of Physics\\
                       National Cheng Kung University\\
                       Tainan, Taiwan\\

\end{center}
\vspace{1cm}
\begin{center} {\large \bf  Abstract} \end{center}
We perform the dimensional reduction of the spacetime of  a stack of  N D3-branes by the ``twist'' identification of a circle to obtain a new Melvin background.   In the near-horizon limit the  background becomes the magnetic-flux deformed $AdS_5 \times S^4$ or $AdS_4 \times S^5$ spacetime.  After analyzing the classical closed string solutions with several angular momenta in different directions of the deformed spacetimes we obtain two string solutions.  The first solution describes a circular closed string located at a fixed value of deformed $AdS_5$ radius while rotating simultaneously in two planes in deformed $AdS_5$ with equal spins $S$.  The second solution describes a string rotating in deformed $S^5$ with two equal angular momenta $J$ in the two rotation planes.  We investigate the small fluctuations therein and show that the magnetic fluxes have inclination to improve the stability of these classical string solutions.

\vspace{2cm}
\begin{flushleft}
E-mail:  whhwung@mail.ncku.edu.tw\\
\end{flushleft}
%%%%%%%%%%%%%%%%%%%%%%%%%%%%%%%
\newpage
\section {Introduction}
The investigation of  the duality between type IIB superstring theory 
in $AdS_5 \times S^5$  and planar limit of $ N=4$ supersymmetric Yang-Mills 
theory [1,2] and extending it to less supersymmetric cases [3-17] may allow us to find  simple  string-theoretic  descriptions  of  various dynamical aspects of gauge theories, from high-energy scattering to confinement.  This AdS/CFT duality is an example of strong coupling - weak  coupling duality as  the large $N$ perturbative expansion in SYM  theory  assumes that  the 't Hooft coupling $\lambda = g^2_{YM} N $ is small while the string perturbative expansion applies for $ \sqrt \lambda = {R^2\over \alpha'} \gg 1$.  For the  BPS-states the  dependence on $\lambda$  is trivial due to the supersymmetry and results can be directly reproduced on the two sides of the duality [2].  

Generalizing AdS/CFT duality to non-BPS  string mode sector can be guided by semiclassical considerations,  as suggested  in [3-6].   In the BMN case [3] one concentrates on a particular ``semiclassical'' [4] sector of  ``near-BPS''  states represented by small closed strings with center of mass  moving along a large circle of $S^5$ with angular momentum $J \gg 1 $.  For the case of  far from BPS,  the investigations [5,6] had found that using the novel multi-spin string states one can  carry out  the  precise test of the  AdS/CFT duality in a  non-BPS sector by comparing the  ${\lambda \over J^2 }\ll 1$  expansion of the {\it classical} string energy  with  the corresponding {\it quantum}  anomalous dimensions in  perturbative SYM theory [7-9]. 

An important step towards the deeper understanding of AdS/CFT correspondence in its less supersymmetric sector was recently given by Lunin and Maldacena [10].    Lunin and Maldacena  demonstrated that certain deformation of the $AdS_{5}\times S^{5}$ background corresponds to a $\beta$-deformation of $N=4$ SYM gauge theory in which the supersymmetry being broken was studied by Leigh and Strassler [11].  Since then there are many studies of AdS/CFT correspondence in the $\beta$-deformation Lunin-Maldacena background [12-17].

The $\beta$-deformed spacetimes considered by Lunin and Maldacena are constructed through TsT transformation by the combination of T-dualities and a shift on the isometries of the 5-sphere part of  $AdS_{5}\times S^{5}$ background.  In this paper we will consider another deformed $AdS_m\times S^n$ background which is obtained by  performing  the dimensional reduction of the spacetime of the N D3-branes through a ``twist'' identification of a circle [18,19].  The new background may be regarded as that with external Melvin magnetic flux [20].   As the near-horizon limit of a stack of N D3-branes becomes $AdS_{5}\times S^{5}$,  the near-horizon limit of the new Melvin background is a magnetic-flux deformed $AdS_5 \times S^4$ or $AdS_4 \times S^5$ spacetime.   For convenience, let us briefly describe the mechanism of  ``twist'' identification in the Kaluza-Klein dimensional reduction. 

 The Melvin spacetime [20] is used to describe a static cylindrically symmetric magnetic flux tube.  The paper of Gibbons and Maeda [18] had generalized this solution to Kaluza-Klein theory by  realizing that it could be obtained from higher-dimensional theory by simply identifying points in a nonstandard way.  For example,  consider the 11-dimensional flat metric in M-theory, which  is written in cylindrically coordinates
$ds^2=-dt^2+ dy_m dy^m +d\rho^2+\rho^2  d\varphi ^2 + dx_{11}^2 $ with the identifications $(t,y_m ,\rho,\varphi, x_{11}) \equiv (t,y_m,\rho,\varphi+2\pi n_1 R B +2\pi n_2, x_{11}+2\pi  n_1R)$ .  The new feature in here is that under a shift of $x_{11}$, one also  shifts ($twist$) $\varphi$ by $2\pi n_1 R B$.   Therefore we can introduce the new coordinate 
$$\tilde\varphi=\varphi - Bx_{11}.\eqno{(1.1)}$$
and rewrite the metric as $ds^2 = - dt^2+ dy_m dy^m +d\rho^2+\rho^2  (d\tilde\varphi+Bdx_{11}) ^2 + dx_{11}^2 $ with the points $(t,y_m ,\rho,d\tilde\varphi, x_{11})$ and $(t,y_m,\rho,d\tilde\varphi+2\pi n_2, x_{11}+2\pi  n_1R)$ identified.  Now recasting the eleven-dimensional metric in the following canonical form  [19]
$$ ds_{11}^2= e^{-2\phi/3}ds_{10}^2+  e^{4\phi/3} (dx_{11}+2 A_\mu dx^\mu )^2 , \eqno{(1.2)} $$
the corresponding ten-dimensional IIA background is then described by
$$ ds_{10}^{2} = \Lambda^{1/2}\left(-dt^2+dy_mdy^m+d\rho^2\right)
  +\Lambda^{-1/2}\rho^2d\tilde{\varphi}^2 ,\eqno{(1.3)} $$
$$  e^{4\phi/3}=\Lambda \equiv 1+\rho^2B^2 ,~~~~ A_{\tilde{\varphi}}=\frac{B\rho^2}{2\Lambda}.  \eqno{(1.4)} $$
In this decomposition into ten-dimensional fields which do not depend on the $x_{11}$, the ten-dimensional Lagrangian density becomes 
$$ L = R- 2 (\nabla \varphi)^2 - e^{2\sqrt 3 \varphi}~ F_{\mu\nu}F^{\mu\nu},\eqno{(1.5)}$$
and the parameter $B$ is the magnetic field defined by $B^2=\frac{1}{2}F_{\mu\nu}F^{\mu\nu}|_{\rho=0}$.  Since the eleven-dimensional spacetime is flat this metric is expected to be an exact solution of the M-theory including higher derivative terms and the Melvin spacetime is therefore an exact solution of M-theory, which can be used to describe the string propagating in the magnetic tube field background.  

  In section II we  apply the above mechanism of  ``twist'' identification in the Kaluza-Klein dimensional reduction to the spacetime of  a stack of  N D3-branes.  In the near-horizon limit we obtain the spacetime of  ${AdS_4}_{(undeformed)} \times S^5_{(deformed)}$ spacetime.  Using the method of Frolov and Tseytlin [5],  we can find a classical closed string which is rotating in deformed $S^5$ with two equal angular momenta $J$ in the two rotation planes.   

In section III, through the same method,  we obtain the spacetime of  $AdS^5_{(deformed)} \times {S^4}_{(undeformed)}$ and find a classical closed string solution which  locates at a fixed value of deformed $AdS_5$ radius while rotating simultaneously in two planes in deformed $AdS_5$ with equal spins $S$.   In both sections we also use the method of Frolov and Tseytlin [5] to investigate the small fluctuations of the solutions therein.  Our results show that the magnetic fluxes have inclination to improve the stability of these classical string solutions.  In section IV we discuss our results.

%%%%%%%%%%%%%%
\section {String in  Magnetic-flux Deformed $AdS_4 \times S^5$}
\subsection {$AdS_5 \times S^5$ and D3-branes}
The spacetime of a stack of $N$ D3-branes in the string frame is described by the metric [21]
$$ ds^2 = H_3^{-1/2} \eta_{\alpha\beta} dx^\alpha dx^\beta + H_3^{1/2} \delta_{ij} dx^i dx^j , \hspace{2.5cm}\eqno{(2.1a)}$$
$$ e^\Phi =1, \hspace{8cm}\eqno{(2.1b)}$$
$$ F_5  = dH_3^{-1}\wedge dx^0\cdot\cdot\cdot \wedge dx^3+ ^\star(dH_3^{-1}\wedge dx^0\cdot\cdot\cdot \wedge dx^3), \eqno{(2.1c)}$$
and 
$$H_3(r) = 1 + \frac{R^4}{r^4}\,,~~~~~ R^4 = 4\pi g_s N \ell_s^4. \eqno{(2.2)}$$
Far from the sources, namely for $r\gg R$, above metric approaches Minkowski space, since $H_3\sim 1$. In the ``near-horizon'' limit, namely the region $r \ll R$,  we can approximate $H_3 \sim \frac{R^4}{r^4}$, and the metric becomes:
$$ ds^2 = \left( \frac{r^2}{R^2}\ \eta_{\alpha\beta} dx^\alpha dx^\beta 
+ \frac{R^2}{r^2} dr^2 \right) + R^2 d\Omega_5^2\,,\eqno{(2.3)}$$
where we have introduced spherical coordinates on the transverse space, and where $d\Omega_5^2$ is the metric on a round five-sphere. We see that the metric is naturally decomposed into two terms, one of which represents a five-sphere of radius $R$.  Using the coordinates $z=\frac{R^2}{r}$, the above metric becomes:
$$ds^2_{} = \frac{R^2}{z^2} \left( \eta_{\alpha\beta} dx^\alpha dx^\beta 
+ dz^2 \right) + R^2 d\Omega_5^2\,,\eqno{(2.4)}$$
in which the metric of five-dimensional Anti-de Sitter space is written in the  ``Poincar\'e coordinates''.   Therefore the geometry resulting from the near-horizon limit of the background generated by $N$ D3-branes is the $AdS_5 \times S^5$, where the radii of the two spaces are both given by $R$.  In the ``cylindrically coordinates'' the metric of unit-radius Anti-de Sitter space can be expressed as
$$ (ds^2)_{AdS_5} = - \cosh^2 \rho \ dt^2 +  d\rho^2 + \sinh^2\rho \ d\Omega_3^{AdS}, ~~~~d\Omega_3^{AdS} \equiv d \theta^2 + sin^2 \theta   d \phi^2 + \cos^2 \theta \ d \varphi^2,  \eqno{(2.5)}$$ 
and the unit-radius  metric on $S^5$ can be described in the following explicit parameterization  
$$ (ds^2)_{S^5}=  d\gamma^2 + cos^2\gamma\ d\varphi_1^2 + sin^2\gamma\ d\Omega^2,~~~~d\Omega^2 \equiv   (d\psi^2 + cos^2\psi\ d\varphi_2^2+ sin^2\psi\ d\varphi_3^2).\eqno{(2.6)}$$ 

\subsection {Deformed $S^5$}
To obtain the magnetic-flux deformed $AdS_4 \times S^5$ we first express  the metric (2.1) as
$$ds_{10}^2 = H_3^{-1/2} \left(- dt^2+ dx^2+dy^2+dz^2 \right) + H_3^{1/2}  \left(dr^2 + r^2\left( d\gamma^2 + cos^2\gamma\ d\varphi_1^2 + sin^2\gamma\ d\Omega^2 \right) \right).\eqno{(2.7)}$$ 
Then, like that in (1.1) we can twist the angle $\varphi_1$ along a compactified coordinate $z$ by the following substituting
$$\varphi_1 \rightarrow \varphi_1 + B z,\eqno{(2.8)}$$ 
and, like that in (1.2), the line element (2.7)  becomes
$$ds_{10}^2 = H_3^{-1/2} \left(- dt^2+ dx^2+dy^2 \right) + H_3^{1/2}  \left(dr^2 + r^2\left( d\gamma^2 + {cos^2\gamma\ d\varphi_1^2\over 1+ B^2H_3~r^2 cos^2\gamma}\ d\varphi_1^2 \right.\right.$$
$$\left.\left.+ sin^2\gamma\ d\Omega^2 \right) \right) + \left(H_3^{-1/2}+ B^2~H_3^{1/2}r^2 cos^2\gamma \right)\left(dz + {B~cos^2\gamma\ d\varphi_1^2\over 1+ B^2H_3~r^2 cos^2\gamma}\right)^2.\eqno{(2.9)}$$ 
Therefore, like that in (1.3) and (1.4), the lower-dimensional theory can be described by
$$ds_{9}^2 = e^{2\Phi/3 }\,H_3^{-1/2}\left[- dt^2+ dx^2+dy^2 + H_3 \left(dr^2 + r^2\left( d\gamma^2 + {cos^2\gamma\ d\varphi_1^2\over 1+ B^2H_3~r^2 cos^2\gamma} + sin^2\gamma\ d\Omega^2 \right) \right)\right].\eqno{(2.10)}$$ 
$$e^{-4\Phi/3} = H_3^{-1/2}+ H_3^{1/2} ~r^2 cos^2\gamma , ~~~~~ A_{\phi_1} ={B~cos^2\gamma\over 2\left(1+ B^2H_3~r^2 cos^2\gamma\right)} .\eqno{(2.11)}$$
In the near-horizon limit $R\ll r$ the line element (2.10) becomes
$$ds_{9}^2 = \sqrt{R^{2} Z^{-2}+B^2R^2cos\gamma^2}\left[{R^2\over Z^2}(- dt^2+ dx^2+dy^2 +dZ^2) +\right.\hspace{2cm} $$
$$\left.\hspace{6cm} R^2\left(d\gamma^2 + {cos^2\gamma\ d\varphi_1^2\over 1+ B^2Z^{2}cos^2\gamma} + sin^2\gamma\ d\Omega^2 \right)\right],\eqno{(2.12)}$$ 
in which we define $Z\equiv R^2/r$.  It is easy to see that the above geometry  is conformal to ${AdS_4}_{(undeformed)} \times S^5_{(deformed)}$ spacetime.   Searching the string solutions in the above  magnetic-flux deformed spacetime is the next work of this section.

\subsection{String Solution in  Magnetic-flux Deformed $S^5$}
The string action we considered could be written in the conformal gauge in terms of the independent  global coordinates $x^m$  
$$ I= - { \sqrt\lambda  \over 4 \pi } \int d^2 \xi \  G_{mn}(x) \partial_a x^m \partial^a  x^n, ~~~~\lambda \equiv  {R^2 \over  \alpha'},\eqno{(2.13)}$$
in which $\xi^a=(\tau,\sigma)$.  In the conformal gauge $\sqrt {-g} g^{ab} = \eta^{ab}= $diag(-1,1), the equations of motion following from the action should be
supplemented by the conformal gauge  constraints
$$ G_{mn}(x) ( \dot  x^m \dot  x^n +  x'^m   x'^n) =0 , \eqno{(2.14a)}$$
$$ G_{mn}(x) \dot x^m   x'^n =0.  \eqno{(2.14b)}$$
Following the method of Frolov and Tseytlin [5],  we now adopt  the ansatz 
$$ t =\kappa t,~~~ \gamma = \gamma(\sigma),~~~ \psi = \psi(\sigma),~~~\varphi_1= \nu \tau,~~~\varphi_2 = \varphi_3= \omega \tau,  \eqno{(2.15)}$$
and, for convenience, take $x = x_0,~ y=y_0,~ R=Z=1$  to find the rotating string solution. 

Substituting the ansatz (2.15) into metric form (2.12) the associated Lagrangian of the action (2.13) is 
$$L = - {\sqrt \lambda\over 4\pi}\sqrt{1+B^2\,cos\gamma(\sigma)}\left[-\kappa^2+{\nu^2cos^2\gamma(\sigma)\over 1+ B^2cos^2\gamma(\sigma)}\, + \omega^2sin^2\gamma(\sigma) - \gamma(\sigma)'^2-\psi(\sigma)'^2 sin^2\gamma(\sigma)  \right],\eqno{(2.16)}$$
As the deformation we used does not change the properties of the  translational isometries of  coordinates $t$, $\varphi_1$, $\varphi_2$ and $\varphi_3$, there are the corresponding four integrals of motion:
$$ E= P_{t}= {\sqrt \lambda} \int^{2\pi}_0 {d \sigma\over  2 \pi}  \sqrt{1+ B^2\,cos^2\gamma(\sigma)}~\partial _0 t \equiv  {\sqrt \lambda} {\cal E},  \eqno{(2.17)}$$
which is the energy of the solution, and  
$$ J_1= P_{\varphi_1}= {\sqrt \lambda} \int^{2\pi}_0
{d \sigma\over  2 \pi}  \sqrt{1+ B^2\,cos^2\gamma(\sigma)}~{cos^2\gamma(\sigma)\over 1+ B^2cos^2\gamma(\sigma)}\, \partial _0\varphi_1 \equiv  {\sqrt \lambda} {\cal J}_1,  \eqno{(2.18a)}$$
$$ J_2= P_{\varphi_2}= {\sqrt \lambda} \int^{2\pi}_0
{d \sigma\over 2 \pi}  \sqrt{1+ B^2\,cos^2\gamma(\sigma)}~sin^2\gamma(\sigma)\,cos^2\psi(\sigma)~\partial _0\varphi_2 \equiv  {\sqrt \lambda} {\cal J}_2,  \eqno{(2.18b)}$$
$$ J_3= P_{\varphi_3}= {\sqrt \lambda} \int^{2\pi}_0
{d \sigma\over  2 \pi}  \sqrt{1+ B^2\,cos^2\gamma(\sigma)}~sin^2\gamma(\sigma)\,sin^2\psi(\sigma)~\, \partial _0\varphi_2 \equiv  {\sqrt \lambda} {\cal J}_3,  \eqno{(2.18c)}$$
which are the angular momenta of the rotating string in the magnetic-flux deformed $S^5$ space.  To find the values of energy and angular momenta we must know the function of $\gamma(\sigma)$ and $\psi(\sigma)$, and have  relations between $\kappa$, $\nu$ and $\omega$.  This can be obtained by solving the equations of  $\gamma(\sigma)$ and $\psi(\sigma)$ associated to the Lagrangian (2.16), and imposing the conformal gauge constraints of (2.14).

The field equations of $\psi(\sigma)$ and $\gamma(\sigma)$ are
$$0= \left( \psi'(\sigma)~sin^2\gamma(\sigma) \right)',\hspace{10.9cm}  \eqno{(2.19)}$$
$$0= 2\left(\sqrt{1+ B^2 cos^2\gamma(\sigma)} \gamma'(\sigma)\right)' 
+b^2 \kappa^2 + 2\nu^2 - {B^2 \nu^2 cos^2\gamma(\sigma)\over 1+ B^2cos^2\gamma(\sigma)}~  - (\omega^2-1)\left[2+B^2(3cos\gamma^2(\sigma)-1)\right], \eqno{(2.20)}$$
respectively.  The eq.(2.19) could be easily solved by setting
$$ \psi(\sigma) = n \sigma,~~~~\gamma(\sigma) =\gamma_0, \eqno{(2.21)}$$
which are the same as those in the undeformed space [5], and we will consider the case of $n=1$ hereafter .  Using (2.21) the field equation of $\psi(\sigma)$ implies
$$\kappa^2 = \left[- {2\over B^2} + {cos^2\gamma_0\over 1+ B^2cos^2\gamma_0}\right] \nu^2~ + {\omega^2-1\over B^2}\left[2+B^2(3cos^2\gamma_0-1)\right]. \eqno{(2.22)}$$
While the conformal gauge constraints (2.14b) is automatically satisfied the another conformal gauge constraints of (2.14a) implies
$$\kappa^2 =  {cos^2\gamma_0\over 1+ B^2cos^2\gamma_0} \nu^2+ (\omega^2+1) ~sin\gamma^2(\sigma),\eqno{(2.23)}$$
in which we have used the relation (2.21). 

   To proceed, let us consider the string solution with of $\nu=0$.   In this case the eq.(2.23) becomes
$$\kappa^2 =  (\omega^2+1) ~sin\gamma^2_0.\eqno{(2.24)}$$
Using the above relation  the eq.(2.22) implies a simple relation
$$\omega^2 =  {1+ B^2 cos^2\gamma_0 \over 1+ B^2 cos2\gamma_0}. \eqno{(2.25)}$$
Therefore, for a given value of $\gamma_0$ we could find the values of $\kappa$ and $\omega$.  Now, the relations (2.17) and (2.18) become
$${\cal E}  = \int^{2\pi}_0 {d \sigma\over  2 \pi}  \sqrt{1+ B^2\,cos^2\gamma_0}~\kappa = \sqrt{1+ B^2\,cos^2\gamma_0}~ \kappa,  \eqno{(2.26)}$$
and  
$${\cal J}_2 =  ~ \omega~\int^{2\pi}_0 {d \sigma\over  2 \pi}  \sqrt{1+ B^2\,cos^2\gamma_0}~sin^2\gamma_0\,cos^2\sigma  =  ~ {\omega\over 2}~ \sqrt{1+ B^2\,cos^2\gamma_0}~sin^2\gamma_0 \equiv {\cal J} ,  \eqno{(2.27a)}$$
$$ J_3= ~ \omega~\int^{2\pi}_0 {d \sigma\over  2 \pi}  \sqrt{1+ B^2\,cos^2\gamma_0}~sin^2\gamma_0\,sin^2\sigma  =  ~ {\omega\over 2}~ \sqrt{1+ B^2\,cos^2\gamma_0}~sin^2\gamma_0 \equiv {\cal J}.  \eqno{(2.27b)}$$
Therefore, with a help of (2.24) and (2.25), the functions ${\cal E}$ and ${\cal J}$ could be expressed as the functions of a single variable $\gamma_0$.  Finally, the relation between the function ${\cal E}$ and ${\cal J}$ could be obtained.  

   However, the relation between ${\cal E}$ and ${\cal J}$ is getting complex while increasing the magnetic flux $B^2$ and we will consider only the case of weak magnetic flux.  To the leading order of $B^2$ we can use (2.27a) to express $cos^2\gamma_0$ as a function of ${\cal J}$.  Substituting this relations into (2.26) we obtain a simple result
$${\cal E} = 2\sqrt {\cal J}+ {1\over 2}\sqrt {\cal J}\left(1- \sqrt {\cal J}\right) B^2+ O(B^4).\eqno{(2.28)}$$
The second term is the corrected energy raised from the Melvin magnetic flux which deforms the $S^5$.  As the string with ${\cal J} \ge 1$ will be a unstable solution, which is investigated in the following subsection, the above result shows that the stable rotating string has a positive corrected energy.

  Note first that the induced metric of the spinning string solution found in this section is flat.  Second, on the near-horizon limit the RR field read form (2.1c) behave as $r^3R^{-4}$, which is $\ll 1$,  and there does not have NS-NS $B_2$ field in our spacetime after the ``twisting'' Kaluza-Klein reduction.  Thus the action (2.13) is a proper one used here. The property also reveals in the section III.

   It is also worth to mention that one should study strings in the 10d background while the reduction spacetime used in (2.12) is indeed 9d.   Phenomenally, a string will sit at 4d spacetime with extra 6d be compactified.  The 9d metric used here is an instructive spacetime adopted to study the effects of a magnetic flux on a spinning string.  It seems that one could apply the mechanism of  the ``twist'' identification to the 11d M-theory with a stack $N$ M-branes and then use some dualities (such as T duality) to find the magnetic-flux deformed $AdS_5 \times  S^5$.   As the classical string solutions we considered  are merely propagating on the subspace of  magnetic-flux deformed $AdS_4 \times  S^5$, we belive that the property we found will not be changed in the magnetic-flux deformed $AdS_5 \times  S^5$.  The problem remains to be clarified.

\subsection{Stability of the String Solution}
The problems of the stability of the rotating string in the $S^5$ had been investigated in detail  in [5] by Frolov and Tseytlin.  The result of appendix A.2 in [5] has shown that the rotating string is stable only if 
$$0 \le \kappa^2 \le {3\over 2}.\eqno{(2.29)}$$
We claim that the above criterion could still be used in the magnetic-flux deformed $S^5$ background, in the case of weak magnetic flux.  Let us see how to arrive the conclusion.   
  
From (2.12) or (2.16) we can see that the magnetic flux $B^2$ only appears in the combination form ``${1+ B^2\,cos^2\gamma(\sigma)}$''.  Thus, during considering  the fluctuation of  the field $\gamma$ we shall replace $\gamma \rightarrow \gamma_0 +\tilde\gamma(t,\sigma)$ in the original Lagrangian, then the combination form could be approximated by
$$1+ B^2\,cos^2(\gamma_0+\tilde\gamma(t,\sigma)) \approx 1+ B^2\,cos^2\gamma_0, \eqno{(2.30)}$$
in the case with  a small value of $B^2$. Thus, the Lagrangian used to investigate the fluctuation field $\gamma$ in the deformed case is equal to that used in the undeformed case, up to a constant value  ``$\sqrt{1+ B^2\,cos^2\gamma_0}$'', after rescaling the field by $\varphi_1 \rightarrow \left(1+ B^2\,cos^2\gamma_0\right)^{-1/2}\varphi_1$.   Therefore, the criterion (2.29) would not be changed under a small magnetic-flux deformation.  

Note that although the criterion (2.29) implies $0 \le {\cal J} \le {3\over 8}$ in the undeformed system,  this criterion of  angular momentum on a stable rotating string will be changed in the deformed system.   This may be seen from (2.24) and (2.25).   The equations tell us that the value $\kappa$ depends on the $\omega$ and $\omega$ depends on the $B^2$.

  After the calculation from the two equations it is shown that
 $$ {\cal J} \approx {1\over 4} \kappa^2 \left[1+ {1\over 2} \left(1-{\kappa^2 \over 2}\right)B^2\right]+ O(B^4).\eqno{(2.31)}$$
Now we can substitute this relation into eq.(2.29) and finally obtain a new result
 $$0 \le {\cal J} \le {3\over 8} + {3\over 64}B^2 +O(B^4).\eqno{(2.32)}$$
which is the stability criterion of the rotating string in the deformed spacetime.  We thus see that magnetic fluxes have inclination to improve the stability of these classical string solutions.

%%%%%%%%%%%%%%%%%%%%%%%%%%%%%%%
\section {String in  Magnetic-flux Deformed $AdS_5 \times S^4$}
In section 2.2 we use the spacetime of a stack of $N$ D3-branes to perform the mechanism of  ``twist'' identification and obtain the magnetic-flux deformed $AdS_5 \times S^4$ after taking the limit of near horizon.   In fact, we can arrive the same deformed spacetime if we use the spacetime of near-horizon limit, i.e. undeformed $AdS_5 \times S^5$ in (2.4),  to perform the mechanism of  ``twist'' identification.   In finding the magnetic-flux deformed $AdS_5 \times S^4$ it is more convenient to adopt the second approach.   

Thus, we will use the near-horizon limit, i.e. undeformed $AdS_5 \times S^5$ described by (2.5) and (2.6) to perform the  ``twist'' identification by the following substituting  
$$\theta \rightarrow \theta + B \varphi_1.  \eqno{(3.1)}$$
Then, after the calculations, the deformed $AdS^5$ spacetime is described by the following metric and magnetic field 
$$ds^2=\sqrt{1+B^2 sinh^2\rho}\left[- \cosh^2\rho dt^2+d\rho^2+{\sinh^2\rho~ d\theta^2\over1+B^2\sinh^2\rho}+\sinh^2\rho\left(sin^2\theta d\phi^2+\cos^2 \theta d\varphi^2\right)\right] \eqno{(3.2)}$$
$$A_\theta =  {B~\sinh^2\rho\over 2\left(1+  B^2\sinh^2\rho\right)},\hspace{9.2cm}\eqno{(3.3)}$$
which will be used to describe the spinning string on the deformed $AdS_5$ at the fixed point $S^5$ with coordinates $\gamma = \varphi_i=\psi =0$, for convenience.

To find the string solution we can follow the method of Frolov and Tseytlin [5] by adopting  the ansatz 
$$ t =\kappa t,~~~ \rho = \rho(\sigma),~~~ \theta = \theta(\sigma),~~~ \phi = \varphi  = \omega \kappa,~~~ \eqno{(3.4)}$$
Following the method in section 2.3 it is easy to check that the solution for the case of 
$$\rho = \rho_0,~~~ \theta = \sigma,\eqno{(3.5)}$$
could satisfy the field equation of  $\rho(\sigma)$ and $\theta(\sigma)$ and the conformal gauge condition if 
$$ \kappa^2 =  sinh^2\rho_0 +{sinh^2\rho_0 \over 1+B^2 sinh^2\rho_0}.\eqno{(3.6)}$$
$$ \omega^2 = \kappa^2 +1. \hspace{3.4cm}\eqno{(3.7)}$$
The above equations are the corresponding equations (2.24) and (2.25) in the deformed $S^5$ space.

The energy and spin in the deformed $AdS^5$ spacetime could be calculated as before and results are
$$E= P_t = \sqrt \lambda~ \sqrt{1+ B^2\,sinh^2\rho_0}~ \kappa \equiv \sqrt \lambda ~{\cal E} ,  \eqno{(3.8a)}$$
and  
$$S_1 = P_{\phi} =  {\omega\over 2}~ \sqrt \lambda~\sqrt{1+ B^2\,sinh^2\rho_0}~sinh^2\rho_0 \equiv \sqrt \lambda ~{\cal S}=S_2 = P_{\varphi},  \eqno{(3.8b)}$$
Therefore, with a help of (3.6) and (3.7), the functions ${\cal E}$ and ${\cal S}$ could be expressed as the functions of a single variable $\gamma_0$.  Finally, the relation between the function ${\cal E}$ and ${\cal S}$ could be obtained.

 The problem of the stability of the above spinning string could be investigated as that in section 2.4.   From eq.(3.2) we can see that the magnetic flux $B^2$ only appear in the combination form ``${1+ B^2\,sinh^2\rho(\sigma)}$''.  Thus, in considering the fluctuation of the field $\rho$ by replacing $\rho\rightarrow \rho_0 +\tilde\rho(t,\sigma)$ in the original Lagrangian, the combination form could be  approximated by 
$$1+ B^2\,sinh^2(\rho_0+\tilde\rho(t,\sigma)) \approx 1+ B^2\,sinh^2\rho_0,\eqno{(3.9)}$$
in the case with  a small value of $B^2$. Thus, the Lagrangian used to investigate the fluctuation in the deformed case is equal to that  in the undeformed case, up to a constant value  ``$\sqrt{1+ B^2\,sinh^2\rho_0}$'' , after rescaling the field by $\theta \rightarrow \left(1+ B^2\,sinh^2\rho_0\right)^{-1/2}\theta$. Therefore, the criterion
$$0 \le\kappa^2 \le {5\over2}.\eqno{(3.10)}$$
which is derived in appendix A.1 in [5] can be applied in the deformed system.   As that in the section 2, although the criterion (3.10) implies $0 \le {\cal S} \le {5\over 8}\sqrt{7\over 2}$  in the undeformed system,  this criterion of a stable spinning string will be changed in the deformed system.   This fact could be easily seen from (3.6) which shows that $\kappa^2$ is the function of $B^2$. 

Now, from the equations (3.6) and (3.7) we can easily express  ${\cal S}$ as a  function of $\kappa^2$, in the case of weak Melvin magnetic flux.  The relation is
$${\cal S} = {1\over 4} \kappa^2 \sqrt {1+\kappa^2}\left[1+{\kappa^2\over2}B^2\right]+O(B^4).\eqno{(3.11)}$$
Substituting this relation to the criterion (3.10) we finally obtain a new result
 $$0 \le {\cal S} \le {5\over 8}\sqrt{7\over2} + {25\over 32}\sqrt{7\over2}B^2 +O(B^4).\eqno{(3.12)}$$
which is the stability criterion of the spinning string in the deformed $AdS^5$ spacetime.  We thus see that magnetic fluxes have inclination to improve the stability of these classical string solutions. 

Finally, it shall mention two works which also investigate the string theory in  an external-field deformed spacetime.  In [16],  the spinning closed string configurations on the Klebanov-Tseytlin (KT) background was studied. The KT spacetime  is a logarithmic deformation of $AdS_5$ with non-trivial NS B-field and it is conjectured to be dual to a non-conformal N = 1 SU(N + M) $\times$ SU(N) gauge theory.   In [17], Gursoy and Nunez studied SL(3, R) deformations of a type IIB background based on D5 branes that is  conjectured to be dual to N = 1 SYM.   They had argued that this deformation of the geometry corresponds to turning on a dipole deformation in the field theory on the D5 branes.  The similar deformations of the geometry that is dual to N = 2 and N = 0 SYM had also been studied. 

%%%%%%%%%%%%%%%%%%%%%%%%%%%%%%%
\section{Conclusion}
The Melvin metric [20] is a solution of  Einstein-Maxwell theory.   It provides us with a curved space-time background in which the superstring theory can be solved exactly [22].   In the Kaluza-Klein spacetime the Melvin solution is a useful metric to investigate the decay of magnetic field [19,23] and the decay of spacetimes, which is related to the closed string tachyon condensation [24,25].   The fluxbranes in the Melvin spacetime have many interesting physical properties as investigated in the resent literatures [26-28]. 

In this paper we use the mechanism of  ``twist'' identification in the Kaluza-Klein dimensional reduction [18] to find the new Melvin spacetime in which the $AdS_5\times S^5$ is deformed by the magnetic flux.  Depending the coordinate of the twisted circle we consider two deformed spacetime: $ {AdS_4}_{(deformed)} \times {S^5}_{(undeformed)}$ and ${AdS_5}_{(undeformed))} \times {S^4}_{(deformed)}$.  For the both deformed spacetime we use the method in [5] to find the string solutions.  Our results indicate that magnetic fluxes have inclination to improve the stability of these classical string solutions.   Note that the supersymmetry is broken by the magnetic field and there will  in general  appear tachyon when the string is in the magnetic-deformed spacetime.  However, as the classical solution studied in this paper corresponds to the spectrum of large quantum number the tachyon will not be shown.   It will be interesting to find the physical string spectrum by following the method in [22] to see more properties of the effects of the magnetic flux on the string when it is in the magnetic-flux deformed $AdS_n \times S^m$ spacetime.

There are also many problems remain to be studied.    First, besides the motions of rotation and spinning the string may also pulsate on the $AdS^5 \times S^5$ [29].  It is interesting to see the effect of the magnetic flux on the spinning pulsating string.  Next, it is known that the  continuum limit of the $SU(2)$ Heisenberg spin chain was shown to reproduce the action describing strings rotating with large angular momentum in an $S^3$ section of $S^5$ [30]. This identification can improve our understanding of the AdS/CFT correspondence.   A natural question here is to find an integrable spin chain system which related to the string non-linear sigma model on the magnetic-flux deformed spacetime.  At first sight, it may be the $SU(2)$ Heisenberg spin chain with external magnetic field.    Finally, it is important to find the Yang-Mills operators corresponding to the spinning strings in the magnetic-flux  deformed spacetimes. The problems are  investigating now. 
\\
\\
\\
\\%%%%%%%%%%%%%%%%%%%%%%%%%%%%%%%
{\Large \bf  References}
\begin{enumerate}
\item J.~M.~Maldacena, ``The large N limit of superconformal field theories and supergravity,'' Adv.\ Theor.\ Math.\ Phys.\  {\bf 2}, 231 (1998) [Int.\ J.\ Theor.\ Phys.\  38 (1999) 1113  [hep-th/9711200];
 S.~S.~Gubser, I.~R.~Klebanov and A.~M.~Polyakov, ``Gauge theory correlators from non-critical string theory,'' Phys.\ Lett.\ B428 (1998) 105 [hep-th/9802109]; E.~Witten, ``Anti-de Sitter space and holography,'' Adv.\ Theor.\ Math.\ Phys.\   2 (1998) 253 [hep-th/9802150].
\item O.~Aharony, S.~S.~Gubser, J.~M.~Maldacena, H.~Ooguri and Y.~Oz, ``Large N field theories, string theory and gravity,'' Phys.\ Rept. 323 (2000) 183 (2000) [hep-th/9905111]; E.~D'Hoker and D.~Z.~Freedman, ``Supersymmetric gauge theories and the AdS/CFT correspondence,''  [hep-th/0201253].
\item D.~Berenstein, J.~M.~Maldacena and H.~Nastase, ``Strings in flat
space and pp waves from N = 4 super Yang Mills'',  JHEP  0204 (2002) 013 [hep-th/0202021].
\item  S.~S.~Gubser, I.~R.~Klebanov and A.~M.~Polyakov, ``A semi-classical limit of the gauge/string correspondence,'' Nucl.\ Phys.\ B636 (2002) 99 [hep-th/0204051]. 
\item S.~Frolov and A.~A.~Tseytlin, ``Multi-spin string solutions in
$AdS_5 \times S^5$,'' Nucl.\ Phys.\ B668 (2003) 77 [hep-th/0304255].
\item S.~Frolov and A.~A.~Tseytlin, ``Quantizing three-spin string
solution in $AdS_5 \times S^5$,'' JHEP 0307 (2003) 016 [hep-th/0306130];
S.A. Frolov, I.Y. Park, A.A. Tseytlin, ``On one-loop correction to energy of spinning strings in $S^5$,''Phys.Rev. D71 (2005) 026006 [hep-th/0408187]; I.Y. Park, A. Tirziu, A.A. Tseytlin, ``Spinning strings in $AdS_5 \times  S^5$: one-loop  correction to energy in SL(2) sector,'' JHEP 0503 (2005) 013 [hep-th/0501203];
N. Beisert, A.A. Tseytlin, K. Zarembo, ``Matching quantum strings to quantum spins: one-loop vs. finite-size corrections,''Nucl.Phys. B715 (2005) 190-210 [hep-th/0502173].
\item S.~Frolov and A.~A.~Tseytlin, ``Rotating string solutions: AdS/CFT duality in non-supersymmetric sectors,'' Phys.\ Lett.\ B570 (2003) 96 [hep-th/0306143].
\item G.~Arutyunov, S.~Frolov, J.~Russo and A.~A.~Tseytlin, ``Spinning strings in $AdS_5 \times S^5$ and integrable systems,'' Nucl.\ Phys.\ B671 (2003) 3 [hep-th/0307191]; G.~Arutyunov, J.~Russo and A.~A.~Tseytlin, ``Spinning strings in $AdS_5 \times S^5$: New integrable system relations,'' Phys.\ Rev.\ D69 (2004) 086009 [hep-th/0311004].
\item A.~A.~Tseytlin, ``Spinning strings and AdS/CFT duality,'' [hep-th/0311139].
\item  O.~Lunin and J.~Maldacena, ``Deforming field theories with U(1) x U(1) global symmetry and their gravity duals,'' JHEP  0505  (2005)  033  [hep-th/0502086]. 
\item R.~G.~Leigh and M.~J.~Strassler, ``Exactly marginal operators and duality in four-dimensional N=1 supersymmetric gauge theory,'' Nucl.\ Phys.\ B447 (1995) 95 [hep-th/9503121].
\item S.~A.~Frolov, R.~Roiban and A.~A.~Tseytlin, ``Gauge - string duality for superconformal deformations of N = 4 super Yang-Mills theory,'' [hep-/0503192];  S.  Frolov, ``Lax Pair for Strings in Lunin-Maldacena Background,'' JHEP 0505 (2005) 069 [ hep-th/0503201].
\item N.P. Bobev, H. Dimov, R.C. Rashkov, ``Semiclassical strings in Lunin-Maldacena background,'' [hep-th/0506063].
\item  S.M. Kuzenko, A.A. Tseytlin, ``Effective action of beta-deformed N=4 SYM theory and AdS/CFT,''  [hep-th/0508098]; J. G. Russo, ``String spectrum of curved string backgrounds obtained by T-duality and shifts of polar angles,'' [hep-th/0508125].
\item  M. Spradlin, T. Takayanagi and A. Volovich, ``String Theory in Beta Deformed Spacetimes,'' [hep-th/0509036];  R. C. Rashkov, K. S. Viswanathan and Y. Yang, ``Generalizations of Lunin-Maldacena transformation on the $AdS_ 5 \times S^5$ background,'' [hep-th/0509058];  S. Ryang,``Rotating Strings with Two Unequal Spins in Lunin-Maldacena Background,''  [hep-th/0509195].
\item I.R. Klebanov, A.A. Tseytlin, ``Gravity Duals of Supersymmetric SU(N) x SU(N+M) Gauge Theories,'' Nucl.Phys. B578 (2000) 123-138 [hep-th/0002159]; 
M.  Schvellinger, ``Spinning and rotating strings for N=1 SYM theory and brane constructions,'' JHEP 0402 (2004) 066 [hep-th/0309161]; Xiao-Jun Wang, ``Spinning Strings on Deformed $AdS_5 \times T^{1,1}$ with NS B-field,'' Phys. Rev. D72 (2005) 086006 [hep-th/0501029].
\item U. Gursoy, C. Nunez, ``Dipole Deformations of N=1 SYM and Supergravity backgrounds with U(1) X U(1) global symmetry,''  Nucl.Phys. B725 (2005) 45-92 [hep-th/0505100].
\item G.~W.~Gibbons and D.~L.~Wiltshire, ``Space-time as a membrane in higher dimensions,'' Nucl.\ Phys.\ B287 (1987) 717 [hep-th/0109093]; G.~W.~Gibbons and K.~Maeda, ``Black holes and membranes in higher dimensional theories with dilaton fields,'' Nucl.\ Phys.\ B298 (1988) 741.
\item F.~Dowker, J.~P.~Gauntlett, D.~A.~Kastor and J.~Traschen, ``The decay of magnetic fields in Kaluza-Klein theory,'' Phys.\ Rev.\ D52 (1995) 6929 [hep-th/9507143].
\item M.A. Melvin, ``Pure magnetic and electric geons,'' Phys. Lett. 8 (1964) 65.
\item C.~G.~Callan, J.~A.~Harvey and A.~Strominger, ``Supersymmetric string solitons,'' [hep-th/9112030]; A.~Dabholkar, G.~W.~Gibbons, J.~A.~Harvey and F.~Ruiz Ruiz, ``Superstrings And Solitons,'' Nucl.\ Phys.\ B  340 (1990) 33; G.T. Horowitz and A.~Strominger, ``Black strings and P-branes,'' Nucl.\ Phys.\ B  360 (1991) 197.
\item  J.~G.~Russo and A.~A.~Tseytlin, ``Exactly solvable string models of curved space-time backgrounds,'' Nucl.\ Phys.\ B449 (1995) 91 [hep-th/9502038]; ``Magnetic flux tube models in superstring theory,'' Nucl.\ Phys.\ B461 (1996) 131 [hep-th/9508068].
\item M.~S.~Costa and M.~Gutperle, ``The Kaluza-Klein Melvin solution in M-theory,'' JHEP 0103 (2001) 027 [hep-th/0012072].
\item  A. Adams, J. Polchinski and E. Silverstein,  ``Don't Panic! Closed String Tachyons in ALE Spacetimes,''  JHEP 0110 (2001) 029 [hep-th/0108075].
\item J.~R.~David, M.~Gutperle, M.~Headrick and S.~Minwalla, ``Closed string tachyon condensation on twisted circles,''  JHEP 0202 (2002) 041 [hep-th/0111212]; S. Minwalla and  T. Takayanagi,  ``Evolution of D-branes Under Closed String Tachyon Condensation,''  JHEP 0309 (2003) 011 [hep-th/0307248];  T.~Takayanagi and T.~Uesugi, ``Orbifolds as Melvin geometry,''   JHEP 0112 (2001) 004 [hep-th/0110099].
\item Wung-Hong Huang, ``Condensation of Tubular D2-branes in Magnetic Field Background, '' Phys.Rev. D70 (2004) 107901 [hep-th/0405192].
\item M.~Gutperle and A.~Strominger, ``Fluxbranes in string theory,'' JHEP 0106 (2001) 035 [hep-th/0104136]; R.~Emparan and M.~Gutperle, ``From p-branes to fluxbranes and back,'' JHEP 0112 (2001) 023 [hep-th/0111177].
\item  M.~S.~Costa, C.~A.~Herdeiro and L.~Cornalba, ``Flux-branes and the dielectric effect in string theory,'' Nucl.\ Phys.\ B619 (2001) 155. [hep-th/0105023]; T.~Takayanagi and T.~Uesugi,  ``D-branes in Melvin background,'' JHEP 0111 (2001) 036 [hep-th/0110200].
\item J.~A.~Minahan, ``Circular semiclassical string solutions on $AdS_5 \times S^5$,'' Nucl.\ Phys.\ B648 (2003) 203 [hep-th/0209047]; J. Engquist, J. A. Minahan and K. Zarembo, ``Yang-Mills Duals for Semiclassical Strings,'' JHEP 0311 (2003) 063 [hep-th/0310188];  M. Alishahiha, A. E. Mosaffa and H. Yavartanoo, ``Multi-spin string solutions in AdS Black Hole and confining backgrounds,''  Nucl.Phys. B686 (2004) 53 [hep-th/0402007];  A. Khan and A. L. Larsen, ``Improved Stabality for Pulsating Multi-spin String Solitons,'' [hep-th/0502063].
\item M.~Kruczenski, ``Spin chains and string theory,'' Phys. Rev. Lett. 93 (2004) 161602. h[ep-th/0311203]; R.~Hernandez and E.~Lopez, ``The SU(3) spin chain sigma model and string theory,'' JHEP 0404 (2004) 052 [hep-th/0403139]; S. Bellucci, P. Y. Caesteill, J. F. Morales and C. Sochichi, ``SL(2) spin chain and spinning strings on $AdS_5 \times S^5$,'' Nucl.\ Phys.\ B707 (2005) 303 [hep-th/0409086];  B. Stefanski, jr. and  A.A. Tseytlin, ``Super spin chain coherent state actions and $AdS_5 \times S^5$ superstring,''  Nucl.Phys. B718 (2005) 83 [hep-th/0503185].
\end{enumerate}
\end{document}